\documentclass[aip,jcp,numerical,reprint]{revtex4-1}
\usepackage[english]{babel}
\usepackage{microtype}
\usepackage{amsmath,amsthm,amssymb,amsfonts}
\usepackage{mathtools}
\usepackage{siunitx}
\usepackage{graphicx}
\usepackage{booktabs}
\usepackage{xr}
\usepackage[breaklinks=true,colorlinks=true,linkcolor=blue,urlcolor=blue,citecolor=blue]{hyperref}

\usepackage[bordercolor=none,backgroundcolor=none,textsize=scriptsize]{todonotes}

\DeclarePairedDelimiter\avg{\langle}{\rangle}

\DeclareMathOperator\const{\mathit{const.}}

\newcommand{\kB}{k_{\text{B}}}

\DeclareSIUnit\molar{\textsc{M}}

\newcommand{\xu}{x^{\text{u}}}
\newcommand{\xw}{x^{\text{w}}}
\newcommand{\ru}{\vec{r}^{\text{u}}}
\newcommand{\rw}{\vec{r}^{\text{w}}}

\begin{document}

\preprint{}

\title{Systematic errors in diffusion coefficients from long-time molecular dynamics simulations at constant pressure}

\author{S\"{o}ren von B\"{u}low}
\author{Jakob T\'{o}mas Bullerjahn}
\affiliation{Department of Theoretical Biophysics, Max Planck Institute of Biophysics, 60438 Frankfurt am Main, Germany}
\author{Gerhard~Hummer}
\email{gerhard.hummer@biophys.mpg.de}
\affiliation{Department of Theoretical Biophysics, Max Planck Institute of Biophysics, 60438 Frankfurt am Main, Germany}
\affiliation{Institute of Biophysics, Goethe University Frankfurt, 60438 Frankfurt am Main, Germany}
\date{\today}

\begin{abstract}
In molecular dynamics simulations under periodic boundary conditions, particle positions are typically wrapped into a reference box.  For diffusion coefficient calculations using the Einstein relation, the particle positions need to be unwrapped.  Here, we show that a widely used heuristic unwrapping scheme is not suitable for long simulations at constant pressure.  Improper accounting for box-volume fluctuations creates, at long times, unphysical trajectories and, in turn, grossly exaggerated diffusion coefficients.  We propose an alternative unwrapping scheme that resolves this issue. At each time step, we add the minimal displacement vector according to periodic boundary conditions for the instantaneous box geometry. Here and in a companion paper [J. Chem. Phys.~\textbf{XXX}, YYYYY (2020)], we apply the new unwrapping scheme to extensive molecular dynamics and Brownian dynamics simulation data.  We provide practitioners with a formula to assess if and by how much earlier results might have been affected by the widely used heuristic unwrapping scheme.  
\end{abstract}


\maketitle

Molecular dynamics (MD) simulations are routinely performed under periodic boundary conditions (PBC).  The particle positions in full space, $\ru \in \mathbb{R}^{3}$, are then wrapped into a reference simulation box, \emph{e.g.}, centered at the origin with $\rw \in [-L/2,L/2)^{3}$ for a cubic box with edge length $L$.  Calculations of observables, such as the mean squared displacement (MSD), require that the saved, wrapped trajectories $\rw(t_i)$ are unwrapped back into full space, $\rw(t_i)\mapsto \ru(t_i)$, in a post-processing step.  MSDs are routinely used for diffusion coefficient estimation via ad-hoc fitting to the Einstein relation, although recent developments show that more accurate results can be retrieved from either a rigorous analysis of the particle displacements\cite{VestergaardBlainey2014} or by properly accounting for MSD correlations.\cite{BullerjahnvonBuelow2020}  The $t_i$ ($i=0,1,\dots$) are the discrete time steps of the saved trajectory, with $\Delta t = t_{i+1} - t_i$ the time-step size if every structure is considered in the unwrapping procedure.  

Software like \texttt{pbctools} in VMD,\cite{GiorginoHenin} \texttt{trjconv} in Gromacs,\cite{AbrahamMurtola2015} or \texttt{cpptraj} in Ambertools\cite{CaseBen-Shalom2018} all rely on a heuristic scheme to unwrap the position of a particle at a given time step $i+1$ by comparing its current wrapped position to its unwrapped position at the previous time step $i$.  The particle is then iteratively translated by an integer number of box edge lengths (independently for each spatial dimension of an orthorhombic simulation box) towards its unwrapped position at time step $i$, until the distance between both unwrapped positions is smaller than half the box edge length.  This scheme is appropriate for simulations performed in the $NVT$ ensemble, \emph{i.e.}, at constant particle number $N$, volume $V$ and temperature $T$, as long as the time-step size is chosen sufficiently short to avoid having the particles move more than half the box edge length within one time step.  

\begin{figure}[t!]
\begin{center}
\includegraphics{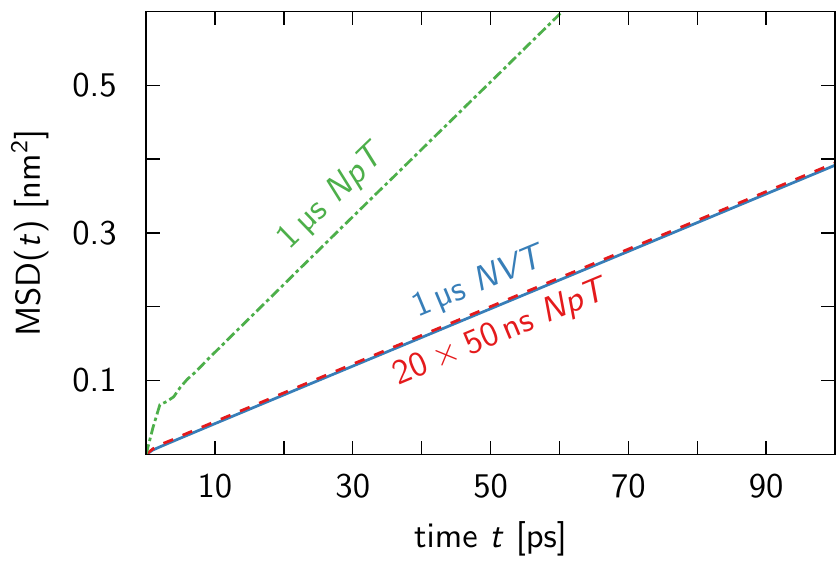}
\caption{Difference in the mean squared displacements of heuristically unwrapped TIP4P-D water trajectories at constant volume and constant pressure.  MD simulations were performed in the $NVT$ and $NpT$ ensembles, respectively, using cubic simulation boxes $(L \approx \SI{2.5}{\nano \meter})$, and unwrapped using Gromacs's \texttt{trjconv} software.  The large discrepancy between the constant-volume (solid blue line) and constant-pressure (dash-dotted green line) simulations highlights the shortcomings of the heuristic unwrapping scheme.  After splitting the $NpT$ trajectory prior to unwrapping into 20 segments (dashed red line), we obtained MSD values comparable to the $NVT$ results. In these \SI{50}{\nano \second} $NpT$ trajectory segments, particles did not diffuse far enough for unphysical unwrapping to occur.
}
\label{fig:mean_squared_displacement}
\end{center}
\end{figure}

\begin{figure*}[t!]
\begin{center}
\includegraphics{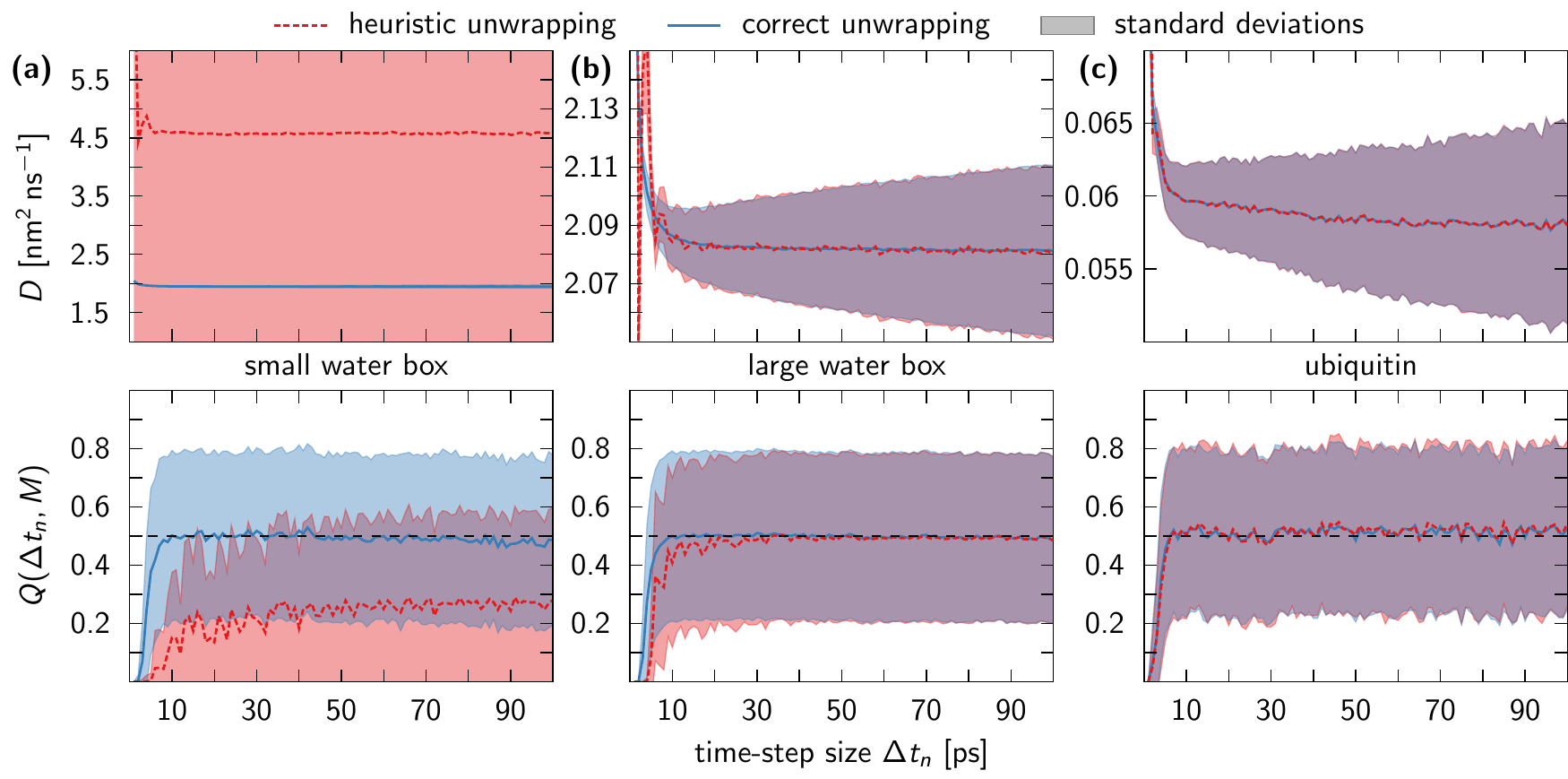}
\caption{Heuristic unwrapping overestimates the diffusion coefficient in long $NpT$ MD simulations of small systems. Top and bottom panels show the estimated diffusion coefficients $D$ and the fit quality factors $Q$, respectively, from MD simulation trajectories of pure TIP4P-D water in (a) a small box (515 water molecules; $L \approx \SI{2.5}{\nano \meter}$) and (b) a large box (4139 water molecules; $L \approx \SI{5}{\nano \meter}$), and (c) of a single ubiquitin molecule in aqueous solution ($L \approx \SI{7.5}{\nano \meter}$). $D$ and $Q$ are shown as functions of the time-step size $\Delta t_{n}$ used to sub-sample the trajectory in the diffusion analysis.\cite{BullerjahnvonBuelow2020} Trajectories were unwrapped at $\Delta t = \SI{1}{\pico \second}$ intervals with the correct (blue) and the heuristic (red) unwrapping schemes according to Eqs.~\eqref{eq:new_scheme} and~\eqref{eq:conventional_scheme}, respectively.  Lines indicate sample averages. Shaded areas represent one sample standard deviation.  We note that in (a) the uncertainty of $D$ from the correct unwrapping scheme is too small to be resolved on the scale of the plot.  }
\label{fig:diffusion_coefficient_quality_factor}
\end{center}
\end{figure*}

In this Communication, we demonstrate that this widely used heuristic trajectory unwrapping scheme is not suitable for simulations at constant pressure $p$. In the $NpT$ ensemble, barostats dynamically adjust the volume of the simulation box to maintain a constant pressure. The particles therefore experience two kinds of displacements: first, their ordinary motion due to collisions and interactions with neighboring particles and, second, a corrective displacement to maintain their relative position inside the box when its volume is varied by the barostat and particle positions are rescaled accordingly.  The heuristic unwrapping scheme fails at constant pressure because it uses the current box size to net-reverse all jumps through the periodic boundaries up to the most recent time step, instead of the respective box sizes for each time step where a jump occurred. Repeated failures then result in unphysical amplifications of the corrective displacements, which eventually dominate over the actual particle motion.  

A consequence of these shortcomings is depicted in Fig.~\ref{fig:mean_squared_displacement}, where we compare MSD estimates from two \SI{1}{\micro \second} MD simulations of TIP4P-D water, which coincide in every aspect except that one was performed in the $NVT$ ensemble, the other in the $NpT$ ensemble (further details of the simulation procedure can be found in the supplementary material).  For a few water molecules in the $NpT$ simulation, the heuristic unwrapping scheme caused an unphysical speed-up, which resulted in an overall acceleration of the average MSD when compared to the $NVT$ simulation.  Importantly, though, the MSD remained linear after \SI{\approx 3}{ps}. The associated diffusion coefficient was also grossly overestimated, as seen in Fig.~\ref{fig:diffusion_coefficient_quality_factor}a, where we compare the heuristically unwrapped $NpT$ data of Fig.~\ref{fig:mean_squared_displacement} to results from a correct unwrapping scheme (see below).  Even without a reference value to compare to, the issues of heuristic unwrapping became apparent in our analysis of the quality factor $Q$, which took values significantly below its expected value of $Q \approx 1/2$ for heuristically unwrapped trajectories (see Fig.~\ref{fig:diffusion_coefficient_quality_factor}a).  The quality factor serves as a measure of how well the data concur with predictions from a minimal diffusive model.\cite{BullerjahnvonBuelow2020} 

\begin{figure*}[t!]
\begin{center}
\includegraphics[width=\textwidth]{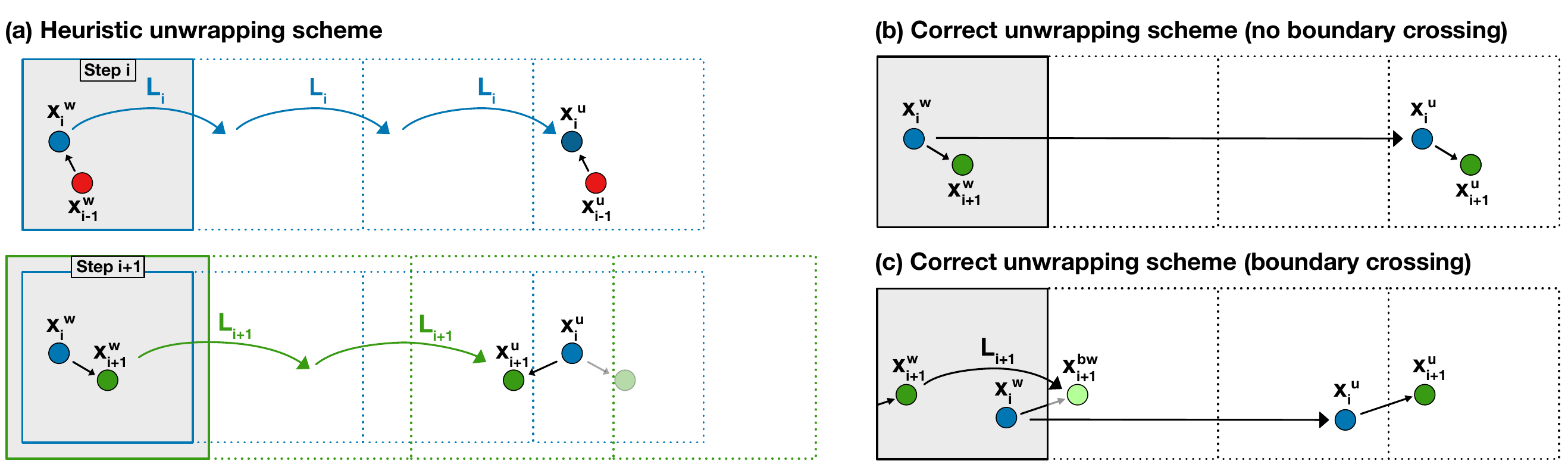}
\caption{Comparison between the heuristic and correct unwrapping scheme.  (a)~In a simulation box (blue) with edge length $L_{i}$ at time step $i$, the heuristic unwrapping scheme [Eq.~\eqref{eq:conventional_scheme}] constructs the unwrapped position $\smash{\xu_{i}}$ from the wrapped coordinate $\smash{\xw_{i}}$ by iterative translation in steps of size $L_{i}$ towards the unwrapped position $\smash{\xu_{i-1}}$ at the previous time step $i-1$, until the condition $\smash{\vert \xu_{i}-\xu_{i-1} \vert} \leq L_{i}/2$ is met.  However, this can lead to artifacts, as demonstrated at time step $i+1$, where the barostat expands the box to length $L_{i+1}$ (green) and the particle is unwrapped into the wrong box, causing it to move left instead of right, which would be the case if the trajectory were correctly unwrapped via Eq.~\eqref{eq:new_scheme} (faint green circle).  (b)~Schematic of the correct unwrapping scheme [Eq.~\eqref{eq:new_scheme}].  In situations, where the particle only diffuses within the simulation box, the unwrapped position $\smash{\xu_{i+1}}$ follows from adding the increment $\smash{\xw_{i+1} - \xw_{i}}$ to $\smash{\xu_{i}}$.  c)~If the particle diffuses out and is placed back into the box according to the PBC, its position $\smash{x_{i+1}^{\text{bw}}}$ before wrapping (``bw'') has to be determined to compute the correct increment to the unwrapped trajectory.  
}
\label{fig:schemes_combined}
\end{center}
\end{figure*}

The effects of heuristic unwrapping can be suppressed in various ways, \emph{e.g.}, by shortening the trajectories drastically (Fig.~\ref{fig:mean_squared_displacement}), increasing the dimensions of the simulation box (Fig.~\ref{fig:diffusion_coefficient_quality_factor}b) or considering molecules that diffuse more slowly (Fig.~\ref{fig:diffusion_coefficient_quality_factor}c).  This is probably the reason why the above-mentioned shortcomings have gone unnoticed for so long.  In the following, we introduce an alternative scheme, sketched in Fig.~\ref{fig:schemes_combined}b, that correctly unwraps trajectories from constant-pressure simulations, and use its output as a reference to quantify the errors introduced by the heuristic unwrapping scheme (Fig.~\ref{fig:schemes_combined}a).  The correct scheme arises naturally when the minimal displacement vector, according to PBC for the instantaneous box geometry, is added to the unwrapped position of the previous time step.  This translates into the following evolution equation for the unwrapped positions $\xu_{i}$ in terms of the wrapped positions $\xw_{i}$ and the box width $L_{i}$, 
\begin{equation}\label{eq:new_scheme}
\xu_{i+1} = \xu_i + (\xw_{i+1} - \xw_{i}) - \left\lfloor \frac{\xw_{i+1} - \xw_{i}}{L_{i+1}} + \frac{1}{2} \right\rfloor L_{i+1} \, , 
\end{equation}
for each spatial dimension in an orthorhombic simulation box.  Here, $\xw_i$ and $\xu_i$ denote the wrapped (``w'') and unwrapped (``u'') one-dimensional coordinates of the particle at time step $i$, respectively. $L_{i}$ is the corresponding box edge length and $\lfloor \cdot \rfloor : \mathbb{R} \to \mathbb{Z}$ is the floor function.  In general, we have triclinic boxes of fluctuating size and shape in $NpT$ simulations.  The simulation box is then defined by the lattice vectors $\smash{\vec{a}_{k = 1,2,3}}$, whose lengths and orientations will fluctuate, with corresponding reciprocal lattice vectors $\smash{\vec{b}_{k}}$ that are obtained by matrix inversion and transposition, $\smash{[\vec{b}_1 \vec{b}_2 \vec{b}_3]^{T}} = \smash{[\vec{a}_1 \vec{a}_2 \vec{a}_3]^{-1}}$.  We generalize Eq.~\eqref{eq:new_scheme} to triclinic boxes by applying PBC to the displacement vector $\smash{\vec{d}_{i+1}^{\text{w}}} = \smash{\rw_{i+1} - \rw_{i}}$, i.e.,
\begin{subequations}\label{eqs:new_scheme_nonortho}
\begin{equation}
\vec{d}_{i+1}^{\text{u}} = \vec{d}_{i+1}^{\text{w}} - [\vec{a}_1 \vec{a}_2 \vec{a}_3] \left\lfloor [\vec{b}_1 \vec{b}_2 \vec{b}_3]^T\vec{d}^{\text{w}}+\left(\begin{array}{c}1/2\\1/2\\1/2\end{array}\right) \right\rfloor \, ,
\end{equation}
and adding the resulting vector to the preceding position of the unwrapped trajectory, 
\begin{equation}
\ru_{i+1}=\ru_{i}+\vec{d}^{\text{u}}_{i+1} \, .  
\end{equation}
\end{subequations}
Here, $\smash{\vec{d}_{i+1}^{\text{u}}}$ is calculated according to the instantaneous box size and shape, and the floor function $\lfloor \cdot \rfloor$ is applied component-wise.  Note that Eqs.~\eqref{eq:new_scheme} and~\eqref{eqs:new_scheme_nonortho} also apply to the wrapped trajectory displacements $\Delta x_{i}^{\text{w}} = x_{i+1}^{\text{w}} - x_{i}^{\text{w}}$ and $\Delta \vec{r}_{i}^{\text{w}} = \vec{r}_{i+1}^{\text{w}} - \vec{r}_{i}^{\text{w}}$, respectively, which must be unwrapped correctly to eliminate the effects of PBC when used as inputs for the covariance-based diffusion coefficient estimator of Ref.~\onlinecite{VestergaardBlainey2014} or for other estimators involving the statistics of particle positions or displacements in full Cartesian space.  

By contrast, for the commonly used heuristic scheme, we have
\begin{equation}\label{eq:conventional_scheme}
\xu_{i+1} = \xw_{i+1} - \left\lfloor \frac{\xw_{i+1} - \xu_{i}}{L_{i+1}} + \frac{1}{2} \right\rfloor L_{i+1} \, .  
\end{equation}
Note that for notational simplicity we concentrate here and in the following on orthorhombic boxes. 
The difference between the unwrapping schemes defined by Eqs.~\eqref{eq:new_scheme} and~\eqref{eq:conventional_scheme} appears to be subtle, boiling down to their respective reference points, namely $\xw_{i}$ and $\xu_i$.  Indeed, we retrieve Eq.~\eqref{eq:conventional_scheme} if $\xw_{i}$ is replaced with $\xu_i$ in Eq.~\eqref{eq:new_scheme}.  Furthermore, the two schemes coincide exactly when applied to simulations in the $NVT$ ensemble, where $L_{i} \equiv L = \const$ holds for all $i$.  

We illustrate the difference between correct and heuristic unwrapping as defined in Eqs.~\eqref{eq:new_scheme} and~\eqref{eq:conventional_scheme}, respectively, by a one-dimensional (1D) Gaussian model.  For this, we consider a Wiener process $\xw$ that evolves on the periodic interval $[-L_{i}/2,L_{i}/2)$.  The boundary positions themselves are realizations of a Gaussian white noise $L_{i}$ (a reasonable assumption, as detailed in the supplementary material), which requires us to constantly rescale the position of our process $\xw$ accordingly.  The wrapped trajectory within the box thus evolves according to
\begin{subequations}\label{eq:gauss_walk}
\begin{gather}
L_{i+1} = \overline{L} + \sigma_L S_{i+1} \, ,
\\
\xw_{i+1} = \frac{L_{i+1}}{L_{i}} \xw_{i} + \sigma_x R_{i+1} - \left\lfloor \frac{\xw_{i}}{L_{i}} + \frac{\sigma_x R_{i+1}}{L_{i+1}} + \frac{1}{2} \right\rfloor L_{i+1} \, , 
\end{gather}
\end{subequations}
where $R$ and $S$ denote uncorrelated normal distributed random variables with zero mean and unit variance, satisfying
\begin{align*}
\avg{R_i} = \avg{S_i} = 0 \, , & & \avg{ R_{i} R_{j} } = \avg{ S_{i} S_{j} } = \delta_{i,j} \, . 
\end{align*}
The Kronecker delta $\delta_{i,j}$ evaluates to one if $i=j$ and zero otherwise.  Typically, the variance $\sigma_{L}^{2}$ of box fluctuations is not specified in MD simulations, but instead the compressibility $\kappa_{T}$ of the system.  Extending our model to three dimensions and assuming isotropic pressure coupling in a cubic box then gives the following approximate relation between the two quantities (see supplementary material),
\begin{equation}
\sigma_{L}^{2} \mathop{\approx}^{\sigma_{L} \ll \overline{L}} \kappa_{T} (9 \beta \overline{L})^{-1} \, .  
\end{equation}
Here, $\beta = (\kB T)^{-1}$ is the inverse thermal energy scale, $T$ the absolute temperature and $\kB$ the Boltzmann constant.  

We generated trajectories for different values of $\smash{\overline{L}}$, $\smash{\sigma_{L}^{2}}$ and $\smash{\sigma_{x}^{2}}$, where the latter coincides (up to a numerical prefactor $2 \Delta t$ for some time-step size $\Delta t$) with the one-dimensional diffusion coefficient $D_{NVT}$ observed in the $NVT$ ensemble.  Each trajectory was unwrapped via both schemes, and for the resulting time series we calculated and fitted the corresponding MSD values using the procedure described in Ref.~\onlinecite{BullerjahnvonBuelow2020} to extract estimates $\sigma^2$ for the effective diffusion coefficient.  Intriguingly, we discovered for our correct unwrapping scheme [Eq.~\eqref{eq:new_scheme}] that box fluctuations do not only add static noise to the MSD, resulting in a constant shift of the MSD curve, but also affect its slope (see supplementary material).  In most practical cases, however, this correction is minuscule and can be neglected.  

\begin{figure}[t!]
\begin{center}
\includegraphics{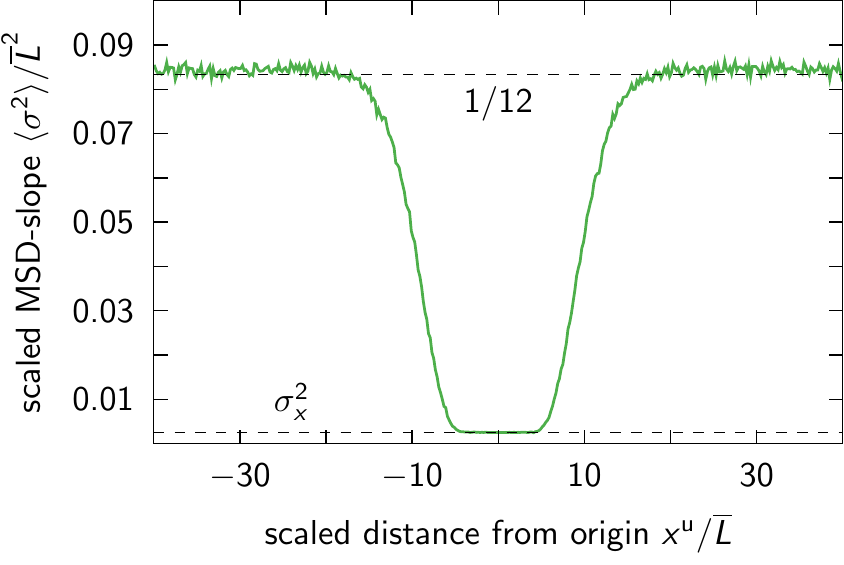}
\caption{Local estimates of the effective diffusion coefficient $\sigma^{2}$ as a function of the particle's unwrapped position $\xu$.  For 1000 heuristically unwrapped trajectories of length $N=\SI{5e5}{}$, we used Eq.~\eqref{eq:local-s2} to determine how the sample average $\avg{\sigma^{2}}$ gradually deviates from its expected value $\sigma_{x}^{2}$ as the particle diffuses further from its starting point.  The wrapped trajectories were generated via Eqs.~\eqref{eq:gauss_walk} using the simulation parameters $\overline{L}=1.0$, $\sigma_{x} = 0.05$ and $\sigma_{L} = 0.02$.  The two horizontal dashed lines indicate the actual diffusion coefficient $\sigma^{2} = \sigma^{2}_{x}$ and the asymptotic limit expected for heuristic unwrapping, $\sigma^{2} = \smash{\overline{L}^{2}} / 12$, respectively.  }
\label{fig:sigma2_heuristic_unwrapping}
\end{center}
\end{figure}

For the heuristic scheme [Eq.~\eqref{eq:conventional_scheme}], we observed a gradual increase in the value of the estimated $\sigma^{2}$ as the particle's unwrapped position deviated further from its origin.  To characterize this effect, we considered very short segments of the unwrapped trajectories, which gave us local estimates 
\begin{equation}\label{eq:local-s2}
\sigma^{2}(\xu_{i}=\xu) = ( \xu_{i+2} - \xu_{i} )^{2} - ( \xu_{i+1} - \xu_{i} )^{2}
\end{equation}
of the diffusion coefficient for every instance where the trajectories reached $\xu$.  Figure~\ref{fig:sigma2_heuristic_unwrapping} shows the sample average over these estimates for bins of width $\smash{\overline{L}/5}$, where $\smash{\sigma^{2}(-5 < \xu < 5)} \approx \sigma_{x}^{2}$.  Further away from the original simulation box $\sigma^{2}$ rises sharply, up to the point where the heuristic unwrapping scheme essentially places the particle randomly inside the interval $\left[ \xu_{i} - L_{i+1}/2, \xu_{i} + L_{i+1}/2 \right)$ at time step $i+1$.  This causes $\sigma^{2}$ to plateau at
\begin{equation*}
\sigma^{2}(\xu \to \pm \infty) \approx \overline{L}^2 / 12 \, ,
\end{equation*}
which coincides with the variance of a uniform distribution on $\left[ \xu_{i} - L_{i+1}/2, \xu_{i} + L_{i+1}/2 \right)$ for all $\xu_{i}$.  Note that the asymptotes of $\sigma^{2}$ are independent of $\sigma_{L}$.  

Figure~\ref{fig:sigma2_heuristic_unwrapping} highlights the fact that errors induced by the heuristic unwrapping scheme remain moderate as long as the particles of interest do not diffuse too far from the original simulation box in the course of the simulation.  The question thus arises whether one can quantify a critical simulation time, beyond which sizable errors in the diffusion coefficient are to be expected?  According to our simulations, this time seems to be on the same order as the average time it takes the heuristic unwrapping scheme to cause a divergent unwrapping event, where the particle is unwrapped into the wrong simulation box, as depicted in Fig.~\ref{fig:schemes_combined}a, for the first time.  We roughly estimated this time with the help of the probability to observe a divergent event at time $i \Delta t$, where $\Delta t$ is the time between consecutive structures to be unwrapped.  As detailed in the supplementary material, we find the following approximate closed-form expression for the critical simulation time,
\begin{equation}\label{eq:critical-time}
t_{\text{crit}} \approx \frac{9 \beta \overline{L}^{5}}{50 \kappa_{T} D_{NVT} [ W_{0}(C^{2/5}) ]^{2}}
\end{equation}
with $C = \smash{9 d N_{\text{p}} \beta \overline{L}^{5} / \left( 25 \sqrt{5 \pi} \kappa_{T} D_{NVT} \Delta t \right)}$.  Here, $d$ denotes the dimension of the simulation box, $N_{\text{p}}$ is the number of diffusing particles of interest in the simulation, and $W_{0}(z)$ is the principal branch of the Lambert $W$ function.  For large arguments, the latter can be replaced with the first two terms of its Taylor expansion, namely
\begin{equation*}
W_{0}(z \gg 1) \approx \ln(z) - \ln \left( \ln(z) \right) + \mathcal{O} \left( \frac{\ln \left( \ln(z) \right)}{\ln(z)} \right) \, .  
\end{equation*}
In general, $D_{NVT}$ is unknown, but practitioners can instead use their estimate of $D_{NpT}$.  This results in a slightly more conservative value for $t_{\text{crit}}$, since $D_{NpT} > D_{NVT}$ for heuristically unwrapped trajectories. For bulk water at ambient conditions, we have experimental values of $D = \SI{2.3}{\nano \meter \squared \per \nano \second}$,\cite{KrynickiGreen1978} $\kappa_{T} = \SI{4.5e-10}{\per \pascal}$,\cite{Kell1970} and a number density of $\smash{N_{\text{p}} / \overline{L}^{3}} \approx \SI{33.3}{\per \nano \meter \cubed}$. We then have to a good approximation $t_{\text{crit}}\approx (\SI{0.0061}{\nano \second}) \smash{N_{\text{p}}^{3/2}} \smash{[\Delta t / (\SI{1}{\pico \second})]^{0.081}}$ for water molecule numbers in the range of $10^{3} < N_{\text{p}} < 10^{6}$ and time-step sizes in the range of $\SI{0.1}{\pico \second} < \Delta t < \SI{10}{\pico \second}$.  For boxes with $N_{\text{p}}=570$, 2900 and 14000 water molecules and $\Delta t=\SI{1}{ps}$, the critical times for water self-diffusion calculations are $t_{\text{crit}} \approx 0.1$, 1 and \SI{10}{\micro \second}, respectively.  

\begin{figure}[t!]
\begin{center}
\includegraphics{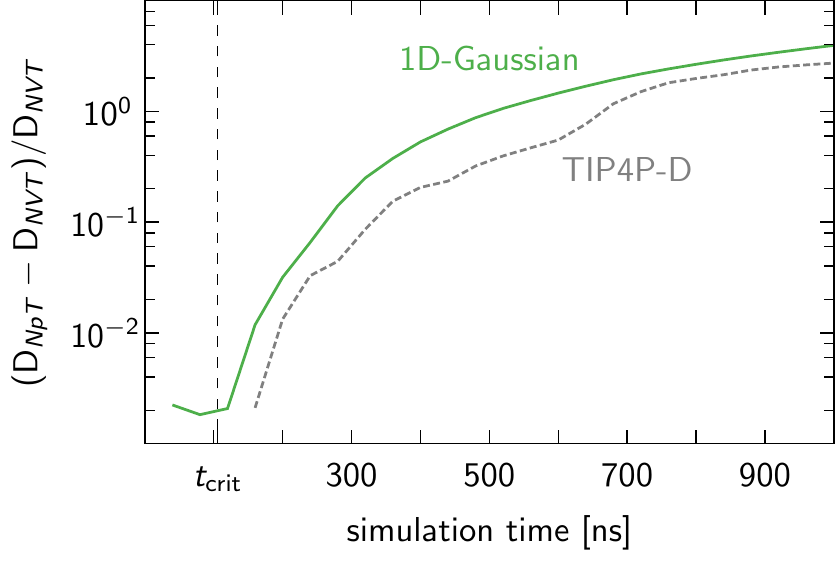}
\caption{Relative error in the diffusion coefficient as a result of heuristic unwrapping depending on the duration of the simulation.  The relative difference between $D_{NpT}$ and $D_{NVT}$ obtained from heuristically unwrapped trajectories is shown for the 1D-Gaussian model (solid green line) and the TIP4P-D water data underlying Figs.~\ref{fig:mean_squared_displacement} and~\ref{fig:diffusion_coefficient_quality_factor}a (gray dashed line).  The diffusion coefficient $D_{NpT}$ was estimated using the formalism of Ref.~\onlinecite{BullerjahnvonBuelow2020} with $\Delta t_{n} = \Delta t_{\text{opt}} = \SI{10}{\pico \second}$ and $M=5$.  Lines indicate sample averages. The 1D-Gaussian model and the critical time $t_{\text{crit}}$ [Eq.~\eqref{eq:critical-time}, black dashed line] were evaluated using the parameters $\smash{\overline{L}} = \SI{2.49}{\nano \meter}$, $\kappa_{T} = \smash{\SI{4.5e-10}{\per \pascal}}$ (corresponding to $\sigma_L = \SI{0.0092}{\nano \meter}$), $D_{NVT} = \smash{\SI{1.95}{\nano\meter\squared \per \nano \second}}$, $d=3$ and $N_{\text{p}} = 515$, which were either directly read off the MD simulation files or extracted from the correctly unwrapped trajectories.  All trajectories were unwrapped using a time-step size of $\Delta t=\SI{2}{\pico \second}$.  At short simulation times ($\lesssim \SI{160}{\nano \second}$), the relative error for the TIP4P-D water data fluctuates around zero and was therefore omitted from the plot.  
}
\label{fig:critical_simulation_time}
\end{center}
\end{figure}

To test our estimate of the critical time $t_{\text{crit}}$ after which we expect the heuristic scheme to fail, we reexamined the heuristically unwrapped TIP4P-D water trajectories from the smaller simulation box (see Fig.~\ref{fig:diffusion_coefficient_quality_factor}a) by truncating them at different points.  We then determined how the resulting effective diffusion coefficient $D_{NpT}$ varies with the length $N$ of the unwrapped trajectory.  In this calculation of $D_{NpT}$, we used a time-step size of $\Delta t_{n} = \Delta t_{\text{opt}} = \SI{10}{\pico \second}$, as described in Ref.~\onlinecite{BullerjahnvonBuelow2020}, to suppress nonlinearities in the MSD on short time scales.  While the trajectories of Figs.~\ref{fig:mean_squared_displacement} and~\ref{fig:diffusion_coefficient_quality_factor} were unwrapped using a time-step size of $\Delta t = \SI{1}{\pico \second}$, we chose here $\Delta t = \SI{2}{\pico \second}$ to suppress the effect of box-fluctuation correlations (see supplementary material), which we do not account for in the 1D-Gaussian model.  Our results are presented in Fig.~\ref{fig:critical_simulation_time}, next to corresponding synthetic three-dimensional data generated by our 1D-Gaussian model [Eq.~\eqref{eq:gauss_walk}].  The critical time, according to Eq.~\eqref{eq:critical-time}, provides a reasonable estimate of the run length, beyond which incorrect heuristic unwrapping events cause significant errors for both the MD data the 1D-Gaussian model.  Further simulations using our 1D-Gaussian model confirm the validity of Eq.~\eqref{eq:critical-time} for a broad range of simulation parameters (see supplementary material).  We therefore advise practitioners to calculate $t_{\text{crit}}$ and compare it to their simulation time if they suspect that the heuristic unwrapping scheme has affected their results in the past.  

In this Communication, we have provided extensive evidence to show that the heuristic unwrapping scheme, as implemented in popular simulation and visualization software, is not appropriate for MD simulations in the $NpT$ ensemble. Initially, its use causes only negligible deviations from the correctly unwrapped trajectory, but, as the simulation progresses, divergent unwrapping events (Fig.~\ref{fig:schemes_combined}a) eventually set in that result in a significant overestimation of diffusion coefficients.  This is especially evident for fast diffusing molecules in small simulation boxes, as seen in our MD simulation of TIP4P-D water (Figs.~\ref{fig:mean_squared_displacement} and \ref{fig:diffusion_coefficient_quality_factor}a).  In the future, the increasing performance of highly parallel GPU architectures will steadily extend the time scales covered by MD simulations, which will increase the chance of noticeable errors in the diffusion coefficient estimated from incorrectly unwrapped trajectories.  By applying PBC on the displacement vector at each time step according to the instantaneous box geometry, the correct unwrapping scheme in Eqs.~\eqref{eq:new_scheme} and \eqref{eqs:new_scheme_nonortho} circumvents the errors arising from the use of the heuristic scheme. In principle, Cartesian-space particle displacements could also be collected ``on the fly'' at the discrete steps of time integration and barostatting, summed for, say, \SI{1}{\pico \second}, and then saved for subsequent analysis. However, trajectory unwrapping not only allows us to reconstruct the aggregate particle displacements with minimal assumptions, but also to (re)analyze trajectories from standard MD codes. Going forward, we urge that the correct unwrapping scheme be implemented in the standard simulation-analysis packages.

\section*{Supplementary material}

See the supplementary material for details on MD simulation procedures, an analysis of box-volume fluctuations in simulations of TIP4P-D water, a numerical and analytic study of the MSD of a correctly unwrapped trajectory, and a detailed derivation of the critical simulation time estimate, along with simulation results to test its validity for various parameter combinations.

\section*{Data availability}
The data that support the findings of this study are available from the corresponding author upon reasonable request.

\begin{acknowledgments}

This research was supported by the Max Planck Society (J.T.B., S.v.B. and G.H.) and the Human Frontier Science Program RGP0026/2017 (S.v.B. and G.H.).  

\end{acknowledgments}

\end{document}


\title{Supplementary material: Systematic errors in diffusion coefficients from long-time molecular dynamics simulations at constant pressure}

\author{S\"{o}ren von B\"{u}low}
\author{Jakob T\'{o}mas Bullerjahn}
\affiliation{Department of Theoretical Biophysics, Max Planck Institute of Biophysics, 60438 Frankfurt am Main, Germany}
\author{Gerhard~Hummer}
\email{gerhard.hummer@biophys.mpg.de}
\affiliation{Department of Theoretical Biophysics, Max Planck Institute of Biophysics, 60438 Frankfurt am Main, Germany}
\affiliation{Institute of Biophysics, Goethe University Frankfurt, 60438 Frankfurt am Main, Germany}

\maketitle

\section{Simulation procedures in the $NVT$- and $NpT$-ensemble}

We performed molecular dynamics (MD) simulations of pure TIP4P-D water \cite{PianaDonchev2015} in cubic simulation boxes of edge length \SI{2.5}{\nano \meter} and \SI{5.0}{\nano \meter} in the $NVT$ and $NpT$ ensemble, resulting in four individual simulations.  In addition, we performed an MD simulation of a single ubiquitin protein (PDB identification code: 1ubq\cite{Vijay-KumarBugg1987}) using the Amber99SB*-ILDN-Q force field\cite{Lindorff-LarsenPiana2010, BestHummer2009, HornakAbel2006, BestdeSancho2012}.  The protein was solvated in TIP4P-D water at a concentration of \SI{150}{\milli \molar} NaCl\cite{JoungCheatham2008} in a cubic simulation box of edge length \SI{7.5}{\nano \meter}.  We used GROMACS 2018.6 \cite{AbrahamMurtola2015} to run the simulations, a time step of \SI{2}{\femto \second}, particle-mesh Ewald electrostatics \cite{DardenYork1993}, and a real-space cutoff of \SI{1.2}{\nano \meter}.  A temperature of \SI{300}{K} was maintained using the velocity-rescaling \cite{BussiDonadio2007} thermostat for all simulations and the Parrinello-Rahman barostat \cite{ParrinelloRahman1981} with isotropic pressure coupling was used for all simulations in the $NpT$ ensemble to maintain a pressure of \SI{1}{bar}.  After initial equilibration for \SI{100}{\pico \second} in the $NVT$ ensemble, all simulations were equilibrated for \SI{5}{\nano \second} in the $NpT$ ensemble followed by production in the $NVT$ or $NpT$ ensemble, respectively.  Trajectory coordinates were saved every $\Delta t=\SI{1}{\pico \second}$.  The total duration of the four water simulations was \SI{1}{\micro \second}, while the ubiquitin simulation ran for \SI{2}{\micro \second}. 

\section{TIP4P-D water box fluctuations}

\subsection{Validating 1D-Gaussian model assumptions}

\begin{figure}[t!]
    \includegraphics[width=\textwidth]{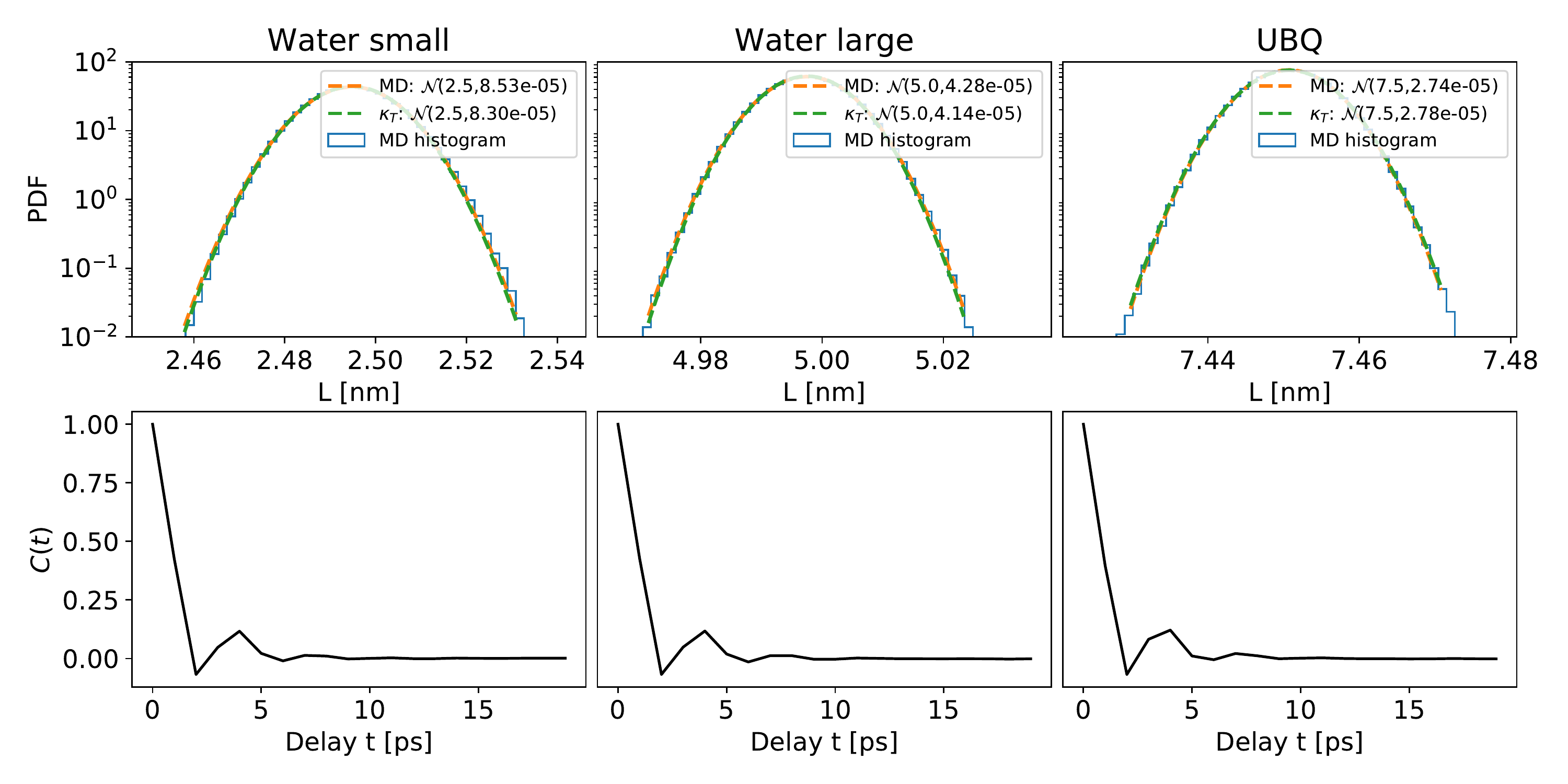}
    \caption{\emph{Top:} Distributions of box edge lengths in molecular dynamics simulations of cubic simulation boxes with isotropic pressure coupling. Blue lines: Histogram from MD simulation. Orange dashed lines: Normal distribution with mean and variance determined from the histogram (blue). Green dashed lines: Expected (normal) distribution of box edge lengths given the compressibility, mean box edge length and temperature. \emph{Bottom:} Normalized autocorrelation functions of box edge length fluctuations.}
    \label{fig:bfluct}
\end{figure}

\noindent
Our 1D-Gaussian model in the main text assumes box fluctuations in constant-pressure simulations to be Gaussian and white.  To justify these assumptions, we examined three molecular dynamics simulations performed at constant pressure in cubic boxes with average edge lengths of $\sim$\SI{2.5}{\nano \meter}, \SI{5.0}{\nano \meter} and \SI{7.5}{\nano \meter}, respectively, as described above.


For each simulation box, we considered every recorded realization $L_{i}$ of the fluctuating box edge lengths and collected them in (normalized) histograms, which are compared to normal distributions $\mathcal{N}(\overline{L}, \sigma_{L}^{2})$ in Fig.~\ref{fig:bfluct}.  On the one hand, the mean $\smash{\overline{L}}$ and variance $\sigma_{L}^{2}$ were computed using the recorded realizations $L_{i}$. On the other hand, we used Eq.~(5) of the main text and specified compressibilities $\kappa_{T}=\SI{4.5e-10}{Pa^{-1}}$ for all simulations to estimate $\sigma_{L}^{2}$. Both approaches give comparable results and fit the histograms nicely for all three box edge lengths, which confirms our assumption that the $L_{i}$ are Gaussian distributed.  

Figure~\ref{fig:bfluct} also depicts the autocorrelation function $C(t)$ of the recorded box fluctuations.  If the $L_{i}$ were truly white, $C(t)$ would resemble a $\delta$-distribution.  However, we observe modest correlations (and anticorrelations) at delay times smaller than \SI{10}{\pico \second}, which can lead to discrepancies between our 1D-Gaussian model and MD simulations, as discussed below.  On longer time scales, the fluctuations are fully uncorrelated and the assumption of white noise in our 1D-Gaussian model becomes justified.  

\subsection{Relation between box fluctuations and compressibility}

\noindent
As pointed out in the main text, $\sigma_{L}^{2}$ is typically not specified for MD simulations, but instead the compressibility $\kappa_{T}$ of the system. In the $NpT$-ensemble, $\kappa_T$ is related to volume fluctuations by~\cite{Hill1986}
\begin{equation*}
\kappa_{T} = \beta \frac{\Avg{(V - \avg{V})^{2}}}{\avg{V}} \, .  
\end{equation*}
Here, $\beta = (\kB T)^{-1}$ is the inverse thermal energy scale, $T$ the absolute temperature and $\kB$ the Boltzmann constant.  For cubic simulation boxes and isotropic pressure coupling, the volume at the time step $i$ is given by $V = \smash{L_{i}^{3}} = \smash{\overline{L}^{3}} (1 + \varepsilon S_{i})^{3}$ with $\varepsilon = \sigma_{L} / \smash{\overline{L}}$ and $S$ denoting a normal distributed random variable with zero mean and unit variance.  In the practical case of $\varepsilon \ll 1$, the first two moments of $V$ read
\begin{align*}
\avg{V} & = \overline{L}^{3} (1 + 3 \varepsilon^{2}) \, ,
\\
\Avg{(V - \avg{V})^{2}} & \approx 9 \overline{L}^{6} \varepsilon^{2} + \mathcal{O}(\varepsilon^{4}) \, ,
\end{align*}
which results in the following expression for the compressibility,
\begin{equation*}
\kappa_{T} \approx 9 \beta \overline{L}^{3} \varepsilon^{2} + \mathcal{O}(\varepsilon^{4}) \, .  
\end{equation*}
By substituting the definition of $\varepsilon$ and solving for $\sigma_{L}^{2}$, one arrives at Eq.~(5) of the main text.  

\section{Mean squared displacement of correctly unwrapped trajectories}

\noindent
To analytically estimate the mean squared displacement (MSD) of the correctly unwrapped trajectory, we need a meaningful expression for $\msd_{i} = \avg{(\xu_{i} - \xu_{0})^{2}}$.  Let us therefore substitute Eq.~(4b) into Eq.~(1) of the main text, giving
\begin{align*}
\xu_{i+1} = \xu_{i} + \bigg[ \frac{L_{i+1}}{L_{i}} - 1 \bigg] \xw_{i} + \sigma_{x} R_{i+1} - \underbrace{\left\lfloor \frac{\xw_{i}}{L_{i}} - \frac{\xw_{i}}{L_{i+1}} + \frac{\sigma_{x} R_{i+1}}{L_{i+1}} + \frac{1}{2} \right\rfloor}_{\approx 0} L_{i+1} \, , 
\end{align*}
where the argument of the floor function is in all practical cases confined to the interval $[0,1)$, thus allowing us to neglect the last term.  The remainder can then be rewritten as follows, 
\begin{equation*}
\xu_{i+1} = \xw_{0} + \sum_{k=0}^{i} \bigg[ \frac{L_{k+1}}{L_{k}} - 1 \bigg] x_{k}^{\text{w}} + \sigma_{X} \sum_{k=1}^{i+1} R_{k} \, ,
\end{equation*}
and the associated MSD is given by
\begin{align}\label{eq:preliminary-msd}
\avg{(\xu_{i} - \xu_{0})^{2}}
\notag
& = \sum_{j,k=0}^{i-1} \bigg\langle \bigg[ \frac{L_{j+1}}{L_{j}} - 1 \bigg] \bigg[ \frac{L_{k+1}}{L_{k}} - 1 \bigg] \xw_{j} \xw_{k} \bigg\rangle
\\
& \mathrel{\phantom{=}} + 2 \sigma_{x} \sum_{j=1}^{i-1} \sum_{k=1}^{j} \bigg\langle \bigg[ \frac{\overline{L}}{L_{j}} - 1 \bigg] \xw_{j} R_{k} \bigg\rangle + i \sigma_{x}^{2} \, .  
\end{align}
Here, we have exploited the facts that i)~$\avg{\xw_{j} R_{k}} = 0$ must hold $\forall k > j$, and ii)~$S_{j+1}$ does not correlate with any of the other components in the second term, thus reducing $\avg{L_{j+1} \xw_{j} R_{k} / L_{j}}$ to $\avg{\overline{L} \xw_{j} R_{k} / L_{j}}$.  Equation~\eqref{eq:preliminary-msd} can be further simplified in the practical case of $\varepsilon = \sigma_{L} / \overline{L} \ll 1$, where the components $L_{i}/L_{i-1}$ get linearized, i.e.,
\begin{align}\label{eq:asymptotic-msd}
\avg{(\xu_{i} - \xu_{0})^{2}}
\notag
& \mathop{\approx}^{\varepsilon \ll 1} i \sigma_{x}^{2} + \varepsilon^{2} \sum_{j,k=0}^{i-1} \avg{ (S_{j+1} - S_{j}) (S_{k+1} - S_{k}) \xw_{j} \xw_{k} } - 2 \sigma_{x} \varepsilon \sum_{j=1}^{i-1} \sum_{k=1}^{j} \avg{  S_{j} \xw_{j} R_{k} }
\\
& \mathrel{\phantom{\mathop{\approx}^{\varepsilon \ll 1}}} + 2 \sigma_{x} \varepsilon^{2} \sum_{j=1}^{i-1} \sum_{k=1}^{j} \avg{ S_{j} S_{j} \xw_{j} R_{k} } + \mathcal{O} ( \varepsilon^{3} ) \, .  
\end{align}
To facilitate an analytic treatment, let us approximate $\xw$ with a rescaled Wiener process, satisfying
\begin{align*}
\xw_{i+1} & \mathop{\approx}^{\mathrel{\phantom{\varepsilon \ll 1}}} \frac{L_{i+1}}{L_{i}} \xw_{i} + \sigma_{x} R_{i+1} \equiv \frac{L_{i+1}}{\overline{L}} \xw_{0} + \sigma_{x} L_{i+1} \sum_{n=1}^{i+1} \frac{R_{n}}{L_{n}} \, , 
\\
\varepsilon^{2} \xw_{i+1} & \mathop{\approx}^{\varepsilon \ll 1} \varepsilon^{2} \Bigg[ \xw_{0} + \sigma_{x} \sum_{n=1}^{i+1} R_{n} \Bigg] + \mathcal{O} ( \varepsilon^{3} ) \, , 
\\
\varepsilon \xw_{i+1} & \mathop{\approx}^{\varepsilon \ll 1} \varepsilon y_{i+1} + \varepsilon \Bigg[ S_{i+1} \xw_{0} + \sigma_{x} \sum_{n=1}^{i+1} (S_{i+1} - S_{n}) R_{n} \Bigg] + \mathcal{O} ( \varepsilon^{3} ) \, .  
\end{align*}
The individual components of Eq.~\eqref{eq:asymptotic-msd} then evaluate to
\begin{gather*}
\avg{S_{i} S_{j} S_{k}} \mathop{=}^{\mathrel{\phantom{\varepsilon \ll 1}}} \avg{S_{i} S_{j} R_{k}} = \avg{S_{i} R_{j} R_{k}} = \avg{R_{i} R_{j} R_{k}} = 0 \, ,
\\
\begin{aligned}
\avg{S_{i} S_{j} S_{k} R_{l}} = \avg{S_{i} R_{j} R_{k} R_{l}} = 0 \, , & & \avg{S_{i} S_{j} R_{k} R_{l}} = \delta_{i,j} \delta_{k,l} \, , 
\end{aligned}
\\
\begin{aligned}
\varepsilon^{2} \avg{ S_{i} S_{j} \xw_{k} \xw_{l} } \mathop{\approx}^{\varepsilon \ll 1} \frac{\sigma_{L}^{2}}{3} \delta_{i,j} + \mathcal{O} ( \varepsilon^{2} ) \, , & & \varepsilon \avg{S_{i} \xw_{j} R_{k}} \mathop{\approx}^{\varepsilon \ll 1} \mathcal{O} ( \varepsilon^{2} ) \, ,
\end{aligned}
\\
\varepsilon^{2} \avg{ S_{i} S_{j} X_{k}^{\text{w}} R_{l} } \mathop{\sim}^{\varepsilon \ll 1} \mathcal{O} ( \varepsilon^{2} ) \, , 
\end{gather*}
which results in the following expression for the MSD,
\begin{equation}\label{eq:analytic-msd}
\avg{(\xu_{i} - \xu_{0})^{2}} \mathop{\approx}^{\varepsilon \ll 1} \frac{2}{3} \sigma_{L}^{2} + i \sigma_{x}^{2} + \mathcal{O} ( \varepsilon^{2} ) \, .  
\end{equation}
However, extensive simulations of Eqs.~(4) of the main text revealed that box fluctuations do not only add static noise to the MSD, as predicted correctly by our analytic calculations, but also affect its slope.  In fact, there is a $\mathcal{O}(\varepsilon)$-term missing in Eq.~\eqref{eq:analytic-msd}, which must arise when the missing $k L_{i+1}$-term in $\xw$, $k \in \mathbb{Z}$, is properly accounted for.  

\begin{figure}[t!]
    \includegraphics[width=0.6\textwidth]{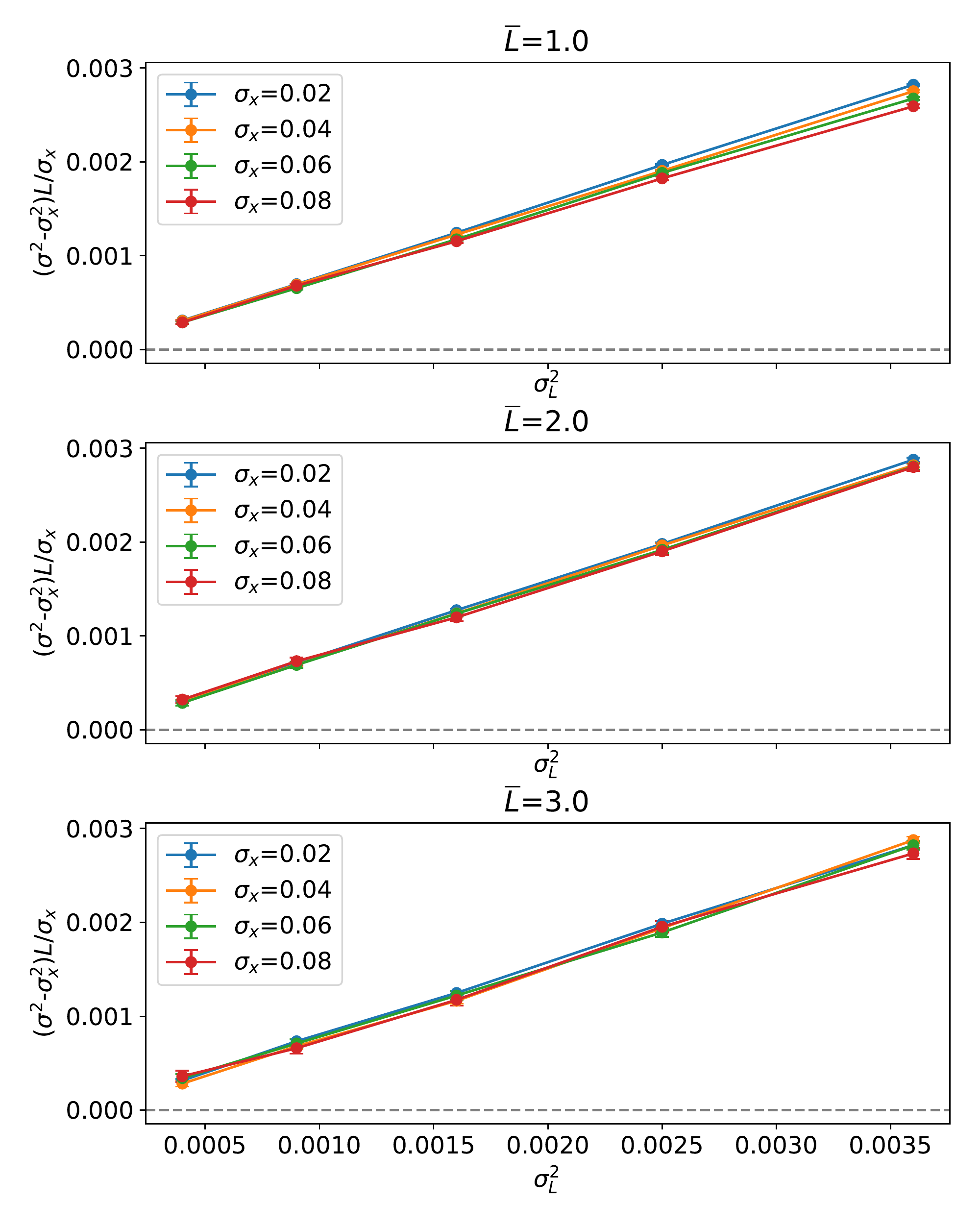}
    \caption{Dependence of $(\sigma^2-\sigma^2_x)\overline{L}/\sigma_x$ on $\sigma^2_L$ from multiple 1D-Gaussian simulations with parameter scans of $\sigma_{L} \in [0.02,0.06]$, $\sigma_{x} \in [0.02,0.08]$, and $\overline{L} \in [1.0,3.0]$. $N=30000$, $\Delta t_{n} = 1$ and $M=5$ for each simulation.}
    \label{fig:alpha}
\end{figure}

To fix this unknown term and its dependence on the system parameters, we performed multiple simulations with $\sigma_{L} \in [0.02,0.06]$, $\sigma_{x} \in [0.02,0.08]$, and $\overline{L} \in [1.0,3.0]$.  The effective diffusion coefficient $\sigma^2$ was determined via generalized least-squares fits to the mean squared displacements using $M=5$ and $\Delta t_{n} = 1$, as described in Ref.~\onlinecite{BullerjahnvonBuelow2020}.  We observed that $(\sigma^{2} - \sigma_{x}^{2}) \overline{L} / \sigma_{x}$ grew linearly with $\sigma_{L}^{2}$ and confirmed our analytic prediction that the intersection point is located at $2 \sigma_{L}^{2}/3$.  The numerical prefactor of the slope correction was determined to be $\alpha \approx 3/4$.  The resulting MSD thus has the functional form
\begin{equation}
\msd_{i} \approx \frac{2}{3} \sigma_{L}^{2} + i \left[ \sigma_{x}^{2} + \frac{3\sigma_{L}^{2}}{4 \overline{L}} \sigma_x \right]
\end{equation}
for moderate values of $\sigma_{x}$.  Figure~\ref{fig:alpha}, top panel, demonstrates that very large input values of $\sigma_{x}$ lead to slight shifts towards smaller values of $\alpha$.  

The effective (one-dimensional) diffusion coefficient in the $NpT$ ensemble thus deviates from its $NVT$ counterpart as follows,
\begin{equation}
D_{NpT} \approx D_{NVT} + \frac{\kappa_{T}}{12 \beta \overline{L}^2 \sqrt{2 \Delta t}} \sqrt{D_{NVT}} \, .  
\end{equation}
Yet, for commonly observed values of $D_{NVT}$ and $\kappa_T$, the spurious term is vanishingly small.  Therefore, if the simulation trajectory is correctly unwrapped, then $D_{NPT} \approx D_{NVT}$ in most practical cases.

\section{Critical simulation time estimation}

\noindent
The derivation of Eq.~(7) in the main text follows from a series of sub-steps: First, we introduce an auxiliary unwrapping scheme, which resembles the heuristic unwrapping scheme of the main text, except that it does not result in divergent events.  This allows us to define conditions that have to be met for a divergent event to occur.  We then proceed to estimate the probability $P(i)$ to observe the first divergent event at the time $i \Delta t$, where $\Delta t$ is the size of the time step for the unwrapped trajectory.  Finally, we derive a closed-form expression for the critical time $t_{\text{crit}} = i_{\text{crit}} \Delta t$, which a simulation should not exceed to suppress artifacts caused by the heuristic unwrapping scheme.  

\subsection{Auxiliary unwrapping scheme}

\noindent
To improve upon the heuristic unwrapping scheme, given by
\begin{align}\label{eq:heuristic-scheme}
\xu_{i+1} = \xw_{i+1} - k_{1} L_{i+1} \, , & & k_{1} = \left\lfloor \frac{\xw_{i+1} - \xu_{i}}{L_{i+1}} + \frac{1}{2} \right\rfloor \, ,
\end{align}
one can rescale each component $\xw_{i}$ of the wrapped trajectory with the corresponding box edge length $L_{i}$.  The rescaled trajectory $\tilde{x}_{i}^{\text{w}} = \xw_{i} / L_{i}$ then evolves inside a box of fixed volume, where the heuristic unwrapping scheme can be safely applied.  Scaling back the unwrapped trajectory with the fluctuating box edge length, i.e.~$\yu = \tilde{x}_{i}^{\text{u}} L_{i}$, then finally results in the following auxiliary unwrapping scheme,
\begin{align}\label{eq:auxiliary-scheme}
\yu_{i+1} = \xw_{i+1} - k_{2} L_{i+1} \, , & & k_{2} = \left\lfloor \frac{\xw_{i+1}}{L_{i+1}} - \frac{\yu_{i}}{L_{i}} + \frac{1}{2} \right\rfloor \, .  
\end{align}
Although the two schemes look very similar, they will eventually start deviating when $k_{1} \neq k_{2}$ for some time step $i$.  

Equations~\eqref{eq:heuristic-scheme} and~\eqref{eq:auxiliary-scheme} imply that $\xu$ and $\yu$ have to satisfy
\begin{align*}
\left\vert \frac{\xu_{i+1}}{L_{i+1}} - \frac{\xu_{i}}{L_{i+1}} \right\vert < \frac{1}{2} \, , & & \left\vert \frac{\yu_{i+1}}{L_{i+1}} - \frac{\yu_{i}}{L_{i}} \right\vert < \frac{1}{2} \, ,
\end{align*}
respectively.  If the two schemes are (still) indistinguishable at time step $i+1$, then $\xu_{j} = \yu_{j}$ $\forall j \leq i+1$.  Furthermore, for $\sigma_{x}^{2} \ll L_{i}$ $\forall i$, $\yu_{i}/L_{i}$ is very close to $\yu_{i+1}/L_{i+1}$, which allows us to (generously) tighten the latter inequality to $\vert \yu_{i+1} / L_{i+1} - \yu_{i} / L_{i} \vert \approx 0$.  The former inequality then reads
\begin{equation*}
\left\vert \frac{\yu_{i}}{L_{i+1}} - \frac{\yu_{i}}{L_{i}} \right\vert < \frac{1}{2}
\end{equation*}
or, in the case of $\varepsilon = \sigma_{L} / \overline{L} \ll 1$,
\begin{equation}\label{eq:inequality}
\vert \yu_{i} \vert \vert S_{i} - S_{i+1} \vert < \overline{L} (2 \varepsilon)^{-1} \, .  
\end{equation}

\subsection{Probability of observing a divergent event}

\noindent
Equation~\eqref{eq:inequality} gives us the following bounds,
\begin{align*}
S_{i+1} \in \mathbb{R} \, , & & S_{i} \in \big( S_{i+1} - \overline{L} (2 \varepsilon \vert \yu_{i} \vert)^{-1}, S_{i+1} + \overline{L} (2 \varepsilon \vert \yu_{i} \vert)^{-1} \big) \, , & & \yu_{i} \in \mathbb{R} \, ,
\end{align*}
which help us define the probability $P(i)$ of the two schemes being indistinguishable at the time step $i+1$.  Said probability is approximately given by
\begin{equation}\label{eq:formal-probability}
P(i) \approx \int_{\mathbb{R}} \ddf{\yu_{i}} p(\yu_{i}) \int_{\mathbb{R}} \ddf{S_{i+1}} p(S_{i+1}) \int_{S_{i+1} - L (2 \varepsilon \vert \yu_{i} \vert)^{-1}}^{S_{i+1} + L (2 \varepsilon \vert \yu_{i} \vert)^{-1}} \ddf{S_{i}} p(S_{i}) \, ,
\end{equation}
where $p(X)$ is the probability distribution function of the process $X$.  According to our model, $p(S_{i})$ and $p(S_{i+1})$ are standard normal distributions, i.e.~$S_{i}, S_{i+1} \sim \mathcal{N}(0,1)$, while $p(\yu_{i})$ is somewhat more complex.  However, to facilitate an analytic treatment, let us assume that $\yu_{i} \sim \mathcal{N}(0,2\sigma_{L}^{2} / 3 + i \sigma_{x}^{2})$.  

The innermost integral of Eq.~\eqref{eq:formal-probability} gives rise to two error functions, which, in combination with the center integral, result in integrals of the type
\begin{equation*}
\int_{\mathbb{R}} \ddf{x} \erf (a x + b) \sqrt{\frac{1}{2\pi\sigma^{2}}} \e^{-(x - \mu)^{2} / 2 \sigma^{2}} = \erf \bigg( \frac{a \mu + b}{\sqrt{1 + 2 a^{2} \sigma^{2}}} \bigg) \, .  
\end{equation*}
Equation~\eqref{eq:formal-probability} therefore reduces to the following integral,
\begin{align*}
P(i) & \approx \frac{1}{2} \int_{\mathbb{R}} \ddf{\yu_{i}} p(\yu_{i}) \int_{\mathbb{R}} \ddf{S_{i+1}} \sqrt{\frac{1}{2\pi}} \e^{-S_{i+1}^{2} / 2} 
\\
& \mathrel{\phantom{=}} \times \Bigg[ \erf \bigg( \frac{S_{i+1}}{\sqrt{2}} + \frac{\overline{L}}{2\sqrt{2} \varepsilon \vert \yu_{i} \vert} \bigg) - \erf \bigg( \frac{S_{i+1}}{\sqrt{2}} - \frac{\overline{L}}{2\sqrt{2} \varepsilon \vert \yu_{i} \vert} \bigg) \Bigg]
\\
& = 2 \int_{0}^{\infty} \ddf{\yu_{i}} \sqrt{\frac{1}{2 \pi (2 \sigma_{L}^{2} / 3 + i \sigma_{x}^{2})}} \e^{-(\yu_{i})^{2} / 2 (2 \sigma_{L}^{2} / 3 + i \sigma_{x}^{2})} \erf \bigg( \frac{\overline{L}}{2\sqrt{2} \varepsilon \yu_{i} } \bigg) \, ,
\end{align*}
which can be treated with Feynman's integration trick, giving
\begin{align*}
\int_{0}^{\infty} \ddf{x} \erf \bigg( \frac{a}{x} \bigg) \e^{-b x^{2}}
& = \int_{0}^{a} \ddf{a'} \int_{0}^{\infty} \ddf{x} \frac{\mathrm{d}}{\mathrm{d}a'} \erf \bigg( \frac{a'}{x} \bigg) \e^{-b x^{2}} = \int_{0}^{a} \ddf{a'} \frac{2 K_{0}(2 a' \sqrt{b})}{\sqrt{\pi}}
\\
& = a \sqrt{\pi} \big[ K_{0}(2 a \sqrt{b}) L_{-1}(2 a \sqrt{b}) + K_{1} (2 a \sqrt{b}) L_{0}(2 a \sqrt{b}) \big] \, ,
\end{align*}
where $K_{\alpha}(x)$ and $L_{\alpha}(x)$ denote the modified Bessel and Struve functions, respectively, with the Bessel functions being of the second kind.  Our sought-after probability~\eqref{eq:formal-probability} thus reads
\begin{equation}
  P(i) \approx A [ K_{0}(A) L_{-1}(A) + K_{1}(A) L_{0}(A) ]
  \label{eq:BesselStruve}
\end{equation}
with $A = \big( 2 \varepsilon^{2} \sqrt{2 / 3 + i \sigma_{x}^{2} / \sigma_{L}^{2}} \big)^{-1}$.  

\subsection{Closed-form expression for the critical simulation time}

\noindent
With the help of $P(i)$, we can now define the probability $P^{*}(i)$ to observe the \emph{first} diverging event at the time step $i+1$ as follows,
\begin{equation}\label{eq:probability}
P^{*}(i) = 1 - \prod_{j=0}^{i} P(j)^{d N_{\text{p}}} \, .  
\end{equation}
Here, we have taken into account that the system is $d$-dimensional and we are considering the trajectories of $N_{\text{p}}$ particles of interest.  We define the critical time $t_{\text{crit}}$ in terms of a time step $i_{\text{crit}} = t_{\text{crit}} / \Delta t$, for which $P^{*}(i_{\text{crit}}-1) \approx 1$ must hold.  On this time scale $P(i)$ still takes sizeable values, which allows us to expand Eq.~\eqref{eq:probability} in terms of the reciprocal probability $\overline{P}(i) = 1 - P(i)$, i.e.,
\begin{align*}
  P^{*}(i_{\text{crit}}-1) = 1 - \prod_{j=0}^{i_{\text{crit}}-1} [ 1 - \overline{P}(j) ]^{d N_{\text{p}}} \mathop{\approx}^{\overline{P}(j) \ll 1} d N_{\text{p}} \sum_{j=0}^{i_{\text{crit}}-1} \overline{P}(j)
  \le d N_{\text{p}} i_{\text{crit}} \overline{P}(i_{\text{crit}}) \, .  
\end{align*}
In the last step, we used the fact that $P(i)$ increases monotonically with $i$. We can thus get an estimate of $t_{\text{crit}}$ by solving
\begin{equation}\label{eq:equation-for-icrit}
1 - P(i_{\text{crit}}) \mathop{=}^{!} \frac{1}{d N_{\text{p}} i_{\text{crit}}} \, .  
\end{equation}
Equation~(7) of the main text is an approximate solution of this relation obtained by using the asymptotic expansions of the modified Bessel and Struve functions in Eq.~\eqref{eq:BesselStruve} for $A \to \infty$ (corresponding to the situation $\varepsilon = \sigma_{L} / \overline{L} \ll 1$) and $i_{\text{crit}} \gg 2 \sigma_{L}^{2} / 3 \sigma_{x}^{2}$.  

\subsection{Testing $t_{\text{crit}}$ prediction}

\begin{figure}
  \includegraphics[width=\textwidth]{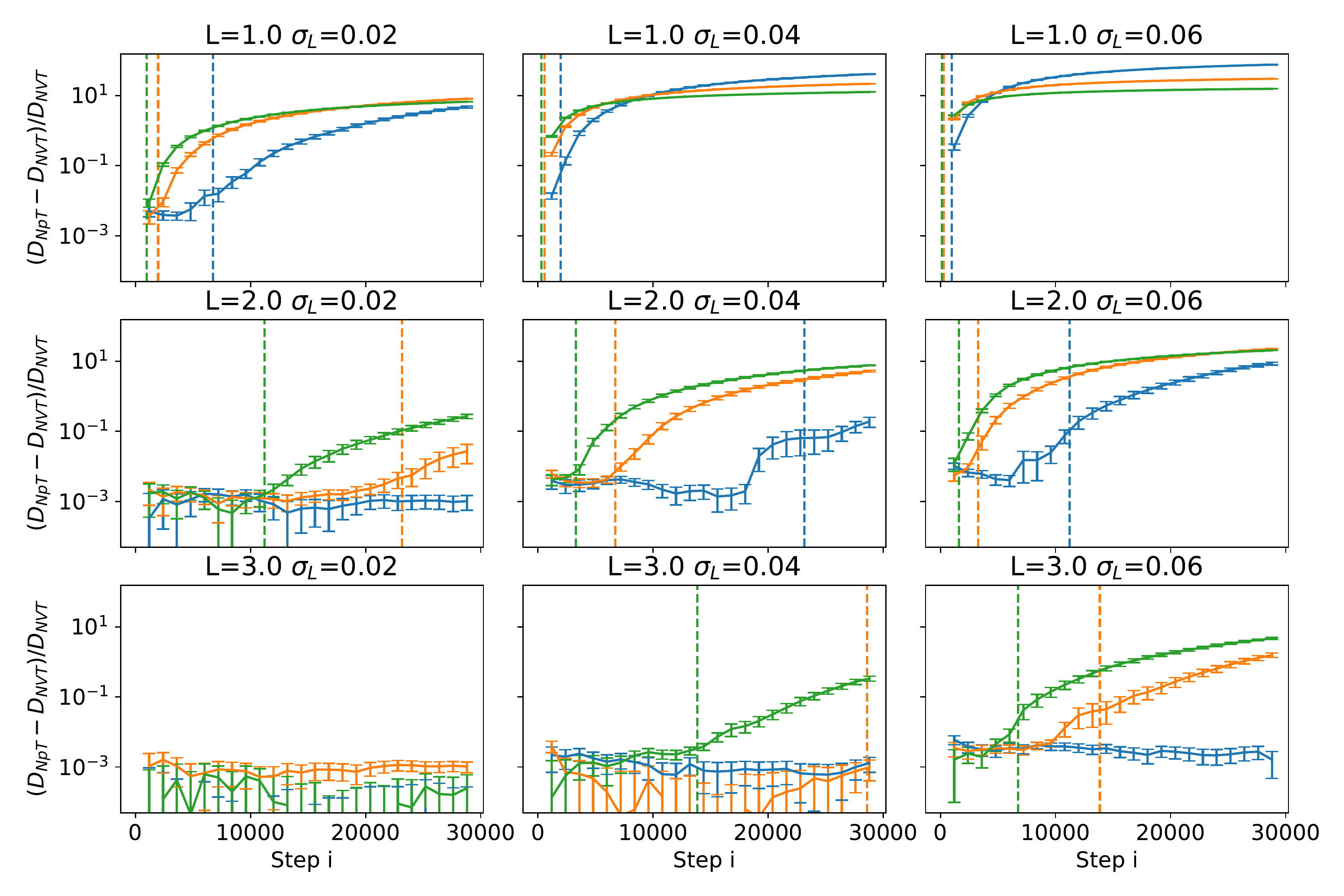}
  \caption{Relative error of diffusion coefficients obtained from heuristic unwrapping, $(D_{NpT}-D_{NVT})/D_{NVT}$, as function of the average box size $\overline{L}$ (panels in vertical direction) and the box-size fluctuations $\sigma_{L}$ (horizontal direction).  The trajectories were obtained from simulations of the 1D-Gaussian model.  The diffusion coefficient $D_{NpT}$ was estimated using the formalism of Ref.~\onlinecite{BullerjahnvonBuelow2020} with $\Delta t_{n}=10$ and $M=5$.  Dotted vertical lines indicate solutions of Eq.~\eqref{eq:equation-for-icrit}, where $N_{\text{p}}$ corresponds to the number of 1D-Gaussian model simulation runs.  Blue: $\sigma_x=0.02$, orange: $\sigma_x=0.04$, green: $\sigma_x=0.06$.}
  \label{fig:s2s_3D}
\end{figure}

We tested if our estimate of the critical time $t_{\text{crit}}$ is applicable over a range of parameters $L$, $\sigma_x$ and $\sigma_L$ of our 1D Gaussian model [Eq.~(4) of the main text]. For this, we evaluated $(D_{NpT}-D_{NVT})/D_{NVT}$ for different simulation times (in analogy to Fig.~5 of the main text) and compared its increase with the prediction $t_{\text{crit}}$ [Eq.~(6) of the main text], as shown in Fig.~\ref{fig:s2s_3D}. In all cases, $(D_{NpT}-D_{NVT})/D_{NVT}$ is small for simulation times shorter than $t_{\text{crit}}$ and becomes sizeable beyond $t_{\text{crit}}$. The critical time estimate is therefore applicable for the range of parameters tested here.  

\section{Suppressing box fluctuation correlations}

\begin{figure}
  \includegraphics[width=0.7\textwidth]{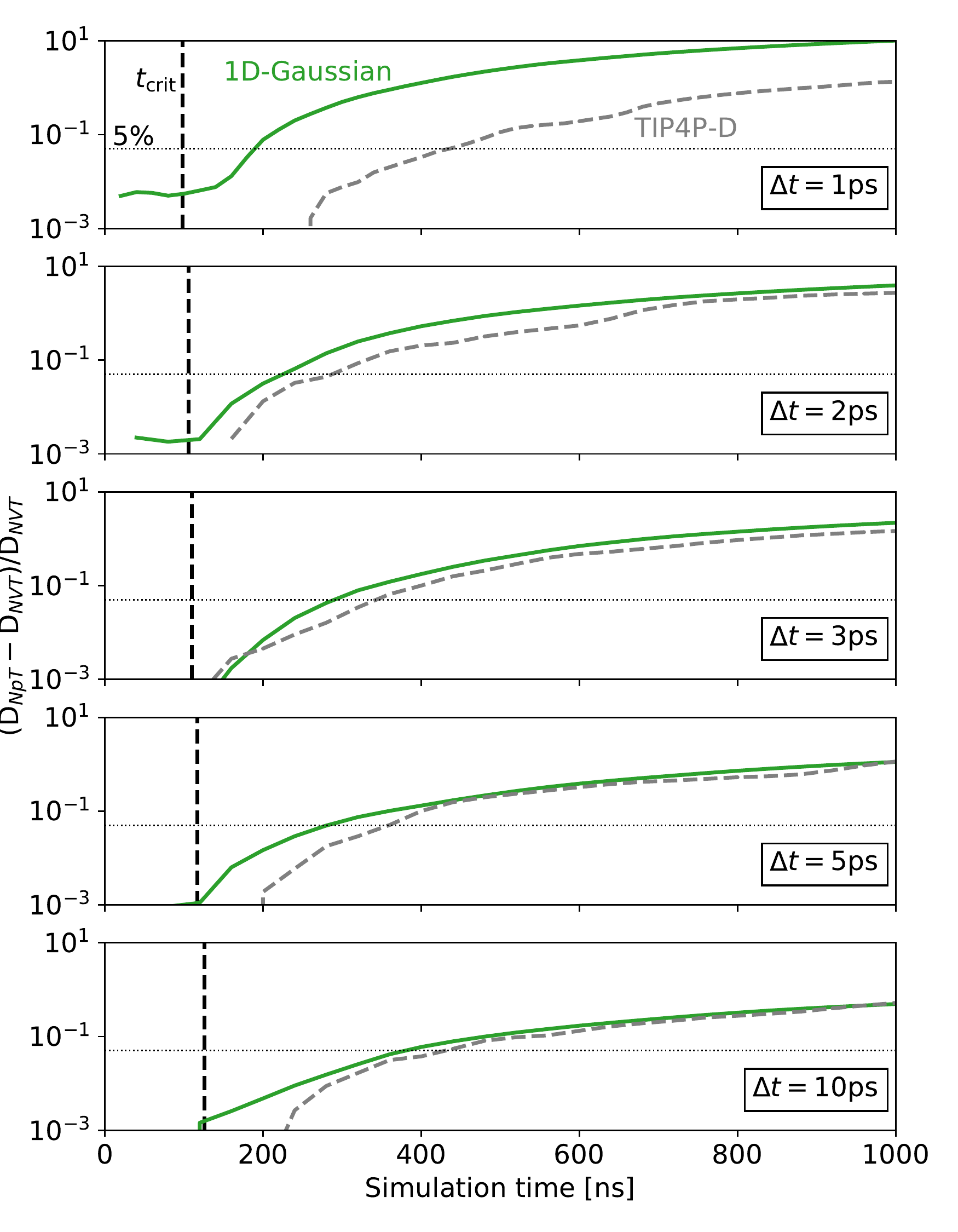}
  \caption{Comparison of relative errors for the 1D-Gaussian model (solid green lines) and the TIP4P-D water data underlying Figs.~1 and 2a of the main text (dashed gray lines) for trajectories saved every 1, 2, 3, 5 or \SI{10}{\pico \second}.  The diffusion coefficient $D_{NpT}$ was estimated using the formalism of Ref.~\onlinecite{BullerjahnvonBuelow2020} with $\Delta t_{n}=\SI{10}{\pico \second}$ and $M=5$.  The curves indicate sample averages. The 1D-Gaussian model and the critical time $t_{\text{crit}}$ [Eq.~(7) of the main text, vertical black dashed lines] were evaluated using the parameters described in Fig.~5 of the main text. The horizontal dotted line serves as a guide to the eye and indicates an error of 5\%. At short simulation times ($\lesssim \SI{160}{\nano \second}$), the relative error for the TIP4P-D water data fluctuates around zero and was therefore omitted from the plot.}
  \label{fig:SI_crit_sim_time}
\end{figure}

In our MD simulations, we observe correlations of the box fluctuations for delay times smaller than \SI{10}{\pico \second} which our 1D-Gaussian model does not account for (Fig. \ref{fig:bfluct}).  The effect is most prominent for the shortest delay of \SI{1}{\pico \second}.  These box fluctuation correlations counteract the error introduced by the heuristic box unwrapping scheme, because $\avg{|L_{i+1}-L_{i}|}$ is smaller for positively correlated box fluctuations than for uncorrelated fluctuations. Therefore, our 1D-Gaussian model and critical time estimate $t_{\text{crit}}$ overestimate the true increase of $D_{NpT}$ for the TIP4P-D water trajectories unwrapped at $\Delta t = \SI{1}{\pico \second}$ (see Fig.~\ref{fig:SI_crit_sim_time}).  When the trajectories are unwrapped at a higher time-step size ($\Delta t \geq \SI{2}{\pico \second}$), the error estimate from 1D-Gaussian model and water data match well.